\documentclass[letter]{aa}  

\usepackage{graphicx}
\usepackage{txfonts}

\begin{document}

\title{Constraining X-ray emission of magnetically arrested disk (MAD) by radio-loud AGNs with extreme ultraviolet (EUV) deficit}

\titlerunning{X-ray emission of MAD}

\author{Shuang-Liang Li
          \inst{1, 2}
          \and
          Minhua Zhou
          \inst{3}
          \and
          Minfeng Gu
          \inst{1}
          }
   \authorrunning{Li et al. }
   
   \institute{Key Laboratory for Research in Galaxies and Cosmology, Shanghai Astronomical Observatory, Chinese Academy of Sciences, 80 Nandan Road, Shanghai 200030, China\\
              \email{lisl@shao.ac.cn; gumf@shao.ac.cn}
             \\
        \and University of Chinese Academy of Sciences, 19A Yuquan Road, 100049 Beijing, PR China\\
        \and 
             School of Physics and Electronic Information, Shangrao Normal University, 401 Zhimin Road, Shangrao 334001, China\\
             \email{zhoumh8@163.com}}

   \date{Received; accepted}

 
  \abstract
   {}
   {Active galactic nuclei (AGNs) with EUV deficit are suggested to be powered by a MAD surrounding the black hole, where the slope of EUV spectra ($\alpha_{\rm EUV}$) is found to possess a well positive relationship with the jet efficiency. In this work, we investigate the properties of X-ray emission in AGNs with EUV deficit for the first time.}
   {We construct a sample of 15 objects with EUV deficit to analyse their X-ray emission. The X-ray luminosity in 13 objects are newly processed by ourself, while the other 2 sources are gathered from archival data.}
   {It is found that the average X-ray flux of AGNs with EUV deficit are 4.5 times larger than that of radio-quiet AGNs (RQAGNs), while the slope of relationship between the optical-UV luminosity ($L_{\rm UV}$) and the X-ray luminosity ($L_{\rm X}$) is found to be similar with that of RQAGNs. For comparison, the average X-ray flux of radio-loud AGNs (RLAGNs) without EUV deficit is about 2-3 times larger than that of RQAGNs. A strong positive correlation between $\alpha_{\rm EUV}$ and radio-loudness ($R_{\rm UV}$) is also reported. However, there is no strong relationship between $L_{\rm X}$ and the radio luminosity ($L_{\rm R}$).}
   {Both the excess of X-ray emission of RLAGNs with EUV deficit and the strong $\alpha_{\rm EUV}$-$R_{\rm UV}$ relationship can be qualitatively explained with MAD scenario, which can help to constrain the theoretical model of MAD.}


   \maketitle

\section{INTRODUCTION}

About 10$\%$ of AGNs are observed to be accompanied by relativistic jets, which have been popular for several decades (see \citealt{2019ARA&A..57..467B} for a review).  There are mainly two jet formation mechanisms in theory, i.e., the so-called Blandford-Znajek (BZ) mechanism \citep{1977MNRAS.179..433B} and the Blandford-Payne (BP) mechanism \citep{1982MNRAS.199..883B}. While BZ and BP mechanisms extract the energy from rotating black hole and accretion disk, respectively, both of them require the presence of large-scale magnetic fields. Large-scale magnetic fields are believed to play a key role on the formation of winds and jets \citep[][etc]{1977MNRAS.179..433B,1982MNRAS.199..883B,1994MNRAS.267..235L,2005ApJ...629..960S,2011MNRAS.418L..79T,2012MNRAS.423.3083M,2013MNRAS.434.1692B, 2019ARA&A..57..467B}, though their origin is still debatable. One promising way is that the field line at outer boundary can be effectively dragged inwards and amplified with the accretion of gas. Usually, the radial velocity in accretion disk will become faster with increasing disk scaled-height \citep{1973A&A....24..337S}.  Therefore, the field line can be easily amplified in a geometrically thick accretion disk due to its shorter advection timescale than diffusive timescale (e.g., \citealt{1994MNRAS.267..235L,2011ApJ...737...94C}), but very difficult for a standard thin disk with very slow radial velocity. However, this problem may be resolved when taking the magnetically driven winds into account, which can improve the radial velocity of gas by transferring lots of angular momentum from accretion disk \citep{2013ApJ...765..149C,2014ApJ...786....6L}.

When more and more magnetic flux is accumulated in the inner region of disk, the magnetic pressure will be comparable or even larger than the gas pressure and thus destroy the symmetry of accretion disk, leading to the generation of a magnetically arrested disk (MAD, e.g., \citealt{2003PASJ...55L..69N}). The magneto-rotational instability (MRI) is suppressed inside MAD, but the gas can still be accreted slowly through magnetic Rayleigh-Taylor (RT) instability (e.g., \citealt{2018MNRAS.478.1837M}). The presence of MAD has been validated by lots of general relativistic magnetohydrodynamic (GRMHD) simulations (e.g., \citealt{2011MNRAS.418L..79T,2012MNRAS.423.3083M,2018MNRAS.480.3547M,2019ApJ...874..168W}). In observations, the EUV emission in some RLAGNs is obviously deficit compared with their radio-quiet counterparts \citep{2002ApJ...565..773T}. This phenomenon had been suggested to be evidence for the presence of MAD \citep[][hereinafter P14 and P15]{2014ApJ...797L..33P,2015ApJ...806...47P}. It is found that the spectral index in EUV band ($\alpha_{\rm EUV}$, $f_{\nu}\sim \nu^{-\alpha_{\rm EUV}}$) has a positive correlation with the jet efficiency ($\eta_{\rm jet}=Q_{\rm jet}/L_{\rm bol}$, where $Q_{\rm jet}$ and $L_{\rm bol}$ are the jet power and bolometric luminosity, respectively), which can be well fitted under the MAD scenario (P14 and P15). The size of MAD ($R_{\rm m}$) given in this scenario is $\sim$ 5.5 gravitational radius ($R_{\rm g}=GM/C^2$) for a modestly rotating black hole. The EUV deficit is caused by the process that a fraction of gravitational energy released between $R_{\rm g}$ and $R_{\rm m}$ may be transferred to jet as Poynting flux by the islands of magnetic flux in MAD (see P14 for details).

While previous study mainly focused on the optical/UV properties of MAD in RLAGNs, for the first time we try to investigate their X-ray properties in this work. The origin of X-ray emission in RLAGNs is still under debate so far, which may come from corona or jet or both. In observations, there is a big difference on the X-ray properties of RQAGNs and RLAGNs. Firstly, the average X-ray flux in RLAGNs is found to be 2-3 times higher than that in RQAGNs (e.g., \citealt{1981ApJ...245..357Z,1987ApJ...323..243W,2021A&A...654A.141L}). Secondly, \citet{1997ApJ...477...93L} reported that RLAGNs have harder 2-10 kev X-ray spectra than RQAGNs by compiling a sample of 23 quasars observed with ROSAT, which is confirmed by \citet{2011ApJS..196....2S} with a larger sample. Comparing the X-ray spectrum of 3CRR quasars and that of radio-quiet quasars, \citet{2020ApJ...893...39Zhou} also gave a similar result. In addition, the X-ray reflection features of RLAGNs are weaker than those of RQAGNs \citep{1998MNRAS.299..449W}. All these results seem to indicate that the contribution of jet to X-ray spectra can't be neglected. However, several recent works suggested a totally different result. For one thing, the slope of $L_{\rm UV}-L_{\rm X}$ is found to be consistent with each other for RLAGNs and RQAGNs \citep{2020MNRAS.496..245Zhu,2021MNRAS.505.1954Z, 2021A&A...654A.141L}. For another, \citet{2018MNRAS.480.2861G,2020MNRAS.492..315G} discovered that the distributions of X-ray photon spectral indices between RLAGNs and their radio-quiet counterpart are very similar (see \citealt{2021MNRAS.505.1954Z} either). This opposite conclusion maybe come from the effect of sample selection. The sample of \citet{2018MNRAS.480.2861G,2020MNRAS.492..315G} is X-ray selected (and optically selected for the sample of \citealt{2021MNRAS.505.1954Z}), which may lead to a result that the radio jet power is very feeble compared to the bolometric luminosity in most of the RLAGNs. These weakly jetted RLAGNs can therefore have different X-ray photon indices comparing with the strong jetted RLAGNs, such as 3CRR quasars of \citet{2020ApJ...893...39Zhou}. P15 also indicated that the weakly jetted RLAGNs have similar $\alpha_{\rm EUV}$ with RQAGNs. However, interestingly, \citet{2005ApJ...635.1203M} demonstrated that both the corona model and the jet model can fit the X-ray data of some Galatic X-ray binaries well and that the jet base may be subsumed to corona in some ways. The 3CRR quasars are low frequency radio selected and have strong jet at large-scale. However, it is still unclear whether all the objects with strong jet harbour MAD, or just being MAD when jet is firstly launching millions of years ago. We focus on the RLAGNs with EUV deficit this work, which should possess MAD on the inner disk region as suggested by P15. The presence of MAD surrounding the black hole may bring remarkable difference to the X-ray emissions, since the structure of disk-corona will greatly change under the case of MAD (e.g., \citealt{2011MNRAS.418L..79T, 2012MNRAS.423.3083M,2019ApJ...874..168W}). In theory, it has been suggested that the X-ray emission will increase when an ADAF (advection-dominated accretion flow) becomes MAD in its inner region \citep{2019ApJ...887..167X}. Nevertheless, how MAD will affect the disk-corona corresponding to the X-ray emission of quasars is still an open issue. This work can constrain future theoretical model for MAD in RLAGNs.

\section{SAMPLE}\label{sample}

\begin{table*}
\normalsize
\flushleft
\caption{The sample.}
\begin{minipage}{\textwidth}
\begin{center}
\begin{tabular}{lllllllllll}
\hline
{Name} & {$z$} & {$\alpha_{\rm {EUV}}$} &  {log$L_{\rm {UV}}$} & {$N_{\rm H}$} & {$\Gamma$} &
{log$L_{\rm X}$} & {$\Delta \log L_{\rm X}$} &  {Telescope} & $\log L_{\rm R}$ & $\log R_{\rm UV}$\\
{} & {} & {} & {[erg s$^{-1}$} & {$\times$ 10$^{20}$ cm$^{-2}$ } & {} & {[erg s$^{-1}$} & {[erg s$^{-1}$} & {} & {[erg s$^{-1}$} & {} \\
{} & {} & {} & {Hz$^{-1}$]} & {} & {} & {Hz$^{-1}$]} & {Hz$^{-1}$]} & {} & {Hz$^{-1}$]} & {}\\
{(1)} &  {(2)} &  {(3)} &  {(4)} &
 {(5)} &  {(6)} & {(7)} &  {(8)} & {(9)}  & {(10)} & {(11)}\\
\hline

1857+566   &  1.6 &  2.87  &  31.10  &   $\leq$45.5 &  1.70$^{+0.5}_{-0.4}$ &  27.95$\pm 1.12$  &  0.81  &    SWIFT &  34.91 & 3.81  \\
1229-021    &  1.05 &  2.65  &  31.14  &   13.14$^{+3.11}_{-3.17}$ &  1.80$^{+0.04}_{-0.06}$   &    27.77$\pm 0.02$  &  0.59  &    Chandra &  34.90 & 3.76    \\
1022+194    &  0.83 &  2.55  &  30.59  &      &  2.55$^{+0.06}_{-0.07}$   &    27.13$\pm 0.08$  &  0.37  &    ROSAT  &  34.39 & 3.80   \\
1040+123    &  1.03 &  2.20  &  30.82  &   $\leq$3.99 &  1.60$^{+0.11}_{-0.06}$   &    27.65$\pm 0.03$  &  0.73  &    Chandra  &  35.11 & 4.30   \\
0743-673   &  1.51  &  2.20   &  31.79  &   $\leq$59.0 &  1.69$^{+0.21}_{-0.15}$   &    28.87$\pm 0.44$  &  1.20  &    SWIFT    &  35.13 & 3.34  \\
1317+520   &  1.06  &  2.18   &  31.32  &   $\leq$1.22 &  1.70$^{+0.04}_{-0.06}$   &    27.54$\pm 0.02$  &  0.23  &    Chandra    &  34.42 & 3.10  \\
1137+660    &  0.65 &  2.0   &  31.18  &   5.24$^{+3.29}_{-3.29}$ &  2.04$^{+0.18}_{-0.18}$   &    27.74$\pm 0.11$  &  0.54  &    Chandra & 33.54 & 2.36   \\
1354+195  &  0.72  &  1.95  &  31.18  &   4.83$^{+1.51}_{-1.75}$ &  1.88$^{+0.04}_{-0.07}$   &    27.85$\pm 0.02$  &  0.65  &    Chandra & 34.58 & 3.40   \\
1340+289  &  0.91  &  1.9  &  30.91  &   $\leq$14.7 &  1.78$^{+0.18}_{-0.12}$   &    28.21$\pm 0.46$  &  1.21  &    SWIFT & 33.92 & 3.01   \\
2340-036    &  0.90 &  1.88  &  31.33  &   $\leq$12.5 &  1.68$^{+0.20}_{-0.08}$   &    28.17$\pm 0.39$  &  0.85  &    SWIFT & 33.91 & 2.58   \\
1241+176    &  1.27 &  1.79  &  31.68  &   $\leq$37.9  &  1.51$^{+0.32}_{-0.18}$    &    27.87$\pm 0.53$   &  0.28  &    SWIFT  & 34.32 & 2.64 \\
1340+606    &  0.96 &  1.78  &  30.80  &   $\leq$6.25 &  1.84$^{+0.09}_{-0.07}$    &    27.26$\pm 0.03$   &  0.34  &    Chandra  & 32.86 & 2.06 \\
0024+224    &  1.12 &  1.77  &  31.55  &     &      &    27.83  &  0.35  &    ROSAT & 34.60 & 3.06  \\
0232-042    &  1.45  &  1.75  &  31.72  &   $\leq$0.82 &  1.87$^{+0.07}_{-0.07}$   &    28.13$\pm 0.03$  &  0.52  &    XMM-Newton & 34.68 & 2.97   \\
0637-752    &  0.65 &  1.75  &  31.06  &   $\leq$0.01 &  1.79$^{+0.04}_{-0.04}$   &    28.16$\pm 0.06$  &  1.04  &    Chandra & 33.67 & 2.61   \\

\hline
\end{tabular}
\end{center}
\end{minipage}
Notes: Col. (1): Source name. Col. (2): Redshift. Col. (3): EUV spectral index. Col. (4): Optical-UV spectral luminosity at 2500 ${\rm} {\AA}$. Col.(5): Intrinsic hydrogen column density. Col. (6): Power-law photon index. Col. (7): X-ray spectral luminosity at 2 keV. Col. (8): Excess of X-ray luminosity in RLAGNs with EUV deficit to that in RQAGNs. Col. (9): X-ray telescope. (10) Radio spectral luminosity at 5 GHz. (11) Radio loudness defined as $R_{\rm {UV}}=L_{\rm {R}}/L_{\rm {UV}}$. \\
\end{table*}

Based on the observations with Hubble Space Telescope (HST), \citet{2002ApJ...565..773T} constructed a sample of 184 quasars to investigate their typical optical-UV spectral properties. The optical/UV continuum of RQAGNs and RLAGNs is found to be similar when the wavelength is longer than 1050$\rm {\AA}$ \citep{1997ApJ...475..469Z, 2002ApJ...565..773T}. However, when the wavelength shortens, $\alpha_{\rm EUV}$ of RLAGNs is found to be larger than that of RQAGNs, where $\alpha_{\rm EUV}$ of RQAGNs and RLAGNs were given as $1.57\pm0.17$ and $1.96\pm0.12$, respectively \citep{2002ApJ...565..773T}. Combining with the jet power, P15 tried to investigate the origin of EUV deficit in RLAGNs and the distribution of magnetic flux tubes in the innermost disk region. All the objects in our sample are compiled from P15. In order to ensure the deficit of EUV emission, we adopt the objects only with $\alpha_{\rm EUV}\geq 1.75$. Fifteen of eighteen objects with $\alpha_{\rm EUV}\geq 1.75$ in P15 are found to be observed in X-ray (see table 1). Among the 15 sources, except for 1022+194 taken from \citet{1997A&A...319..413B} and 0024+224 derived from NED\footnote{http://ned.ipac.caltech.edu/}, the X-ray flux of other 13 sources are gotten by ourselves.

Our X-ray data process of Chandra and XMM-Newton are described in \citet{2020ApJ...893...39Zhou,2021RAA....21....4Z}, and here we provide only an overview. 

All Chandra data were processed with Chandra Interactive Analysis of Observations software (CIAO) v4.12 and Chandra Calibration Database (CALDB) version 4.9.2.1 with CIAO threads\footnote{http://cxc.harvard.edu/ciao/threads/index.html}. We reprocessed Chandra archived data with \textit{chandra\_repro} script, checked background flares, filtered energy between 0.3 and 7.0 $\rm keV$, and inspected piled-up effect with same method in \citet{2020ApJ...893...39Zhou}. The source spectrum was finally extracted from source and background regions at source-centered circle of radius 2.5" and 20-30" annulus, respectively. 

The original XMM-Newton X-ray data were reprocessed with XMM-Newton Scientific Analysis Software (SAS) package. For better X-ray quality, we prefer to use pn dataset which were processed with $epproc$ in SAS-18.0.0 and were filtered in energy range of $0.2-15.0\rm ~ keV$. We filtered out large flares time interval and checked pile-up effect. The source and background X-ray spectrum were then extracted from source-centered radius of 32" and source-free region of 40" around the object. Both Chandra and XMM-Newton X-ray spectra were fitted with absorbed power-law model with fixed Galactic absorption ($phabs*zphabs*powerlaw$). Based on the fitting results of power-law parameters, we estimated the X-ray flux at $2~\rm keV$, in Table 1.

We processed Swift X-ray spectra with the on-line XRT $\&$ UVOT data analysis tool\footnote{https://www.swift.ac.uk/user\_objects/}. With a series of input settings, source name, redshift, 'All' observations, and some default settings, the on-line analysis tool will present the time-averaged X-ray spectrum for all observations and the corresponding power-law fitting results. Thus the flux at 2 keV thus can be calculated.

All the data in columns (1)-(3) are directly taken from P15. The spectral luminosity at 2500${\rm \AA}$ ($L_{\rm UV}$) in column (4) are derived from $L_{\rm EUV}$ at 1100${\rm \AA}$ with a spectral index $\alpha_{\rm o}=-0.5$ ($f_\nu=\nu^{\alpha_{\rm o}}$, e.g., \citealt{2011ApJS..196....2S}), where $L_{\rm EUV}$ at 1100${\rm \AA}$ can be achieved from P15. Columns (5) and (6) are the hydrogen column density and photon index adopted to fit the X-ray data, respectively. $\Delta \log L_{\rm X}$ in column (8) represent the excess of X-ray luminosity observed and the X-ray flux calculated with the $L_{\rm UV}$-$L_{\rm X}$ relationship given by \citet{2010A&A...512A..34L} (see section 3 for details). The radio core luminosity at 5 GHz (column 10) are compiled from NED. For sources without direct observations at 5 GHz, the luminosity was derived from the neighbouring frequencies with a radio spectral index 
index $\alpha_{\rm R} = -0.5$ ($f_\nu \sim \nu^{\alpha_{\rm R}}$; e.g. \citealt{2011ApJS..196....2S}).

\section{RESULTS}\label{results}

\begin{figure}
\centering
\includegraphics[width=8cm]{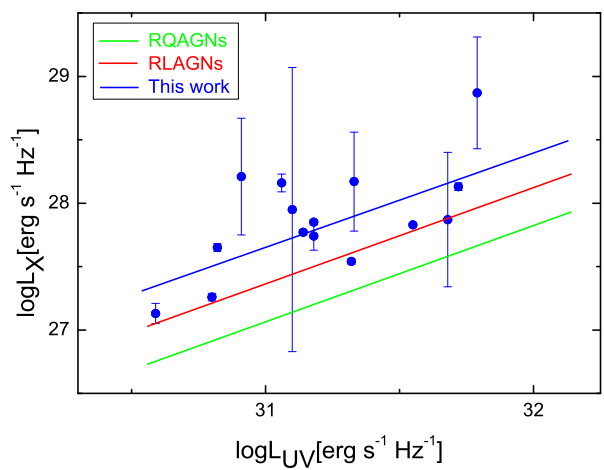}
\caption{The relationship between the X-ray luminosity $L_{\rm X}$ and optical/UV luminosity $L_{\rm UV}$ in RLAGNs with EUV deficit, where the blue line represents our best-fitting result. The green line is the best-fitting result for RQAGNs given by \citet{2010A&A...512A..34L}. The red line corresponds to the relationship for RLAGNs without EUV deficit. }\label{f1}
\end{figure}

\begin{figure}
\centering
\includegraphics[width=8cm]{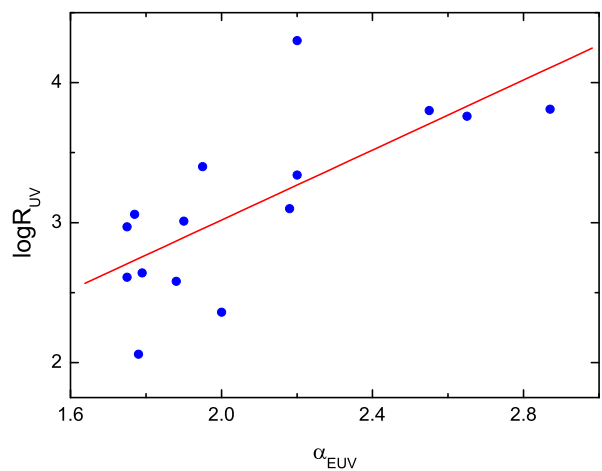}
\caption{The relationship between $R_{\rm UV}$ and $\alpha_{\rm EUV}$, where the red line represents the best-fitting result.}\label{f2}
\end{figure}

\begin{figure}
\centering
\includegraphics[width=8cm]{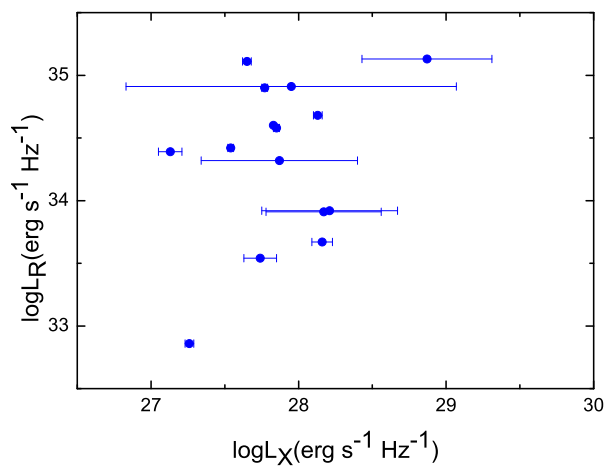}
\caption{The relationship between the radio spectral luminosity $L_{\rm R}$ and the X-ray spectral luminosity $L_{\rm X}$.}\label{f3}
\end{figure}

By constructing a type 1 AGN sample of over 500 sources from XMM-COSMOS survey, \citet{2010A&A...512A..34L} investigated the optical-to-X-ray properties of AGNs and reported a tight positive relationship between X-ray luminosity and optical/UV luminosity in RQAGNs, i.e., $\log L_{\rm X}=0.76\log L_{\rm UV}+3.508$. This relationship can be qualitatively fitted with the classical disk-corona model, where the X-ray flux comes from the inverse Compton scattering of hot corona to the seed photon from an optically thick and geometrically thin accretion disk \citep{2017A&A...602A..79L,2018MNRAS.477..210Q}. A similar relationship has also been achieved in RLAGNs recently \citep{2020MNRAS.496..245Zhu,2021A&A...654A.141L}. However, the X-ray flux in RLAGNs is found to be 2-3 times larger than that in RQAGNs \citep{1981ApJ...245..357Z,1987ApJ...323..243W,2018MNRAS.480.2861G,2021A&A...654A.141L}, though their slope is quite similar. 

In Fig. \ref{f1}, we investigate the correlation between $L_{\rm X}$ at 2 kev and $L_{\rm UV}$ at 2500$\rm \AA$ for the RLAGNs with EUV deficit, where the blue line represents our best-fitting result and the green line is the best-fitting result for RQAGNs given by \citet{2010A&A...512A..34L}.  The red line shows the $L_{\rm UV}$-$L_{\rm X}$ relationship for RLAGNs without EUV deficit, where the X-ray luminosity is 2 times (0.3 dex) larger than the green line \citep{1981ApJ...245..357Z,1987ApJ...323..243W,2018MNRAS.480.2861G,2021A&A...654A.141L}. We find that the X-ray emission in RLAGNs with EUV deficit is, on average, 4.5 and 2.2 times higher than that in RQAGNs and RLAGNs without EUV deficit, respectively. The relationship between $L_{\rm X}$ and $L_{\rm UV}$ can be given as:
\begin{equation}
\log L_{\rm X}=(0.75\pm 0.25) \log L_{\rm UV}+4.40\pm 7.79,
\end{equation}
where the confidence level based on a Pearson test is about 99.0\%. 

$\alpha_{\rm EUV}$ is found to be positively correlated with the jet efficiency $\eta_{\rm jet}$, which can be understood with the MAD scenario in P15. The presence of MAD indicates stronger magnetic field strength in the region surrounding the black hole, which can improve both $\alpha_{\rm EUV}$ and the jet power. We further investigate the relationship between $\alpha_{\rm EUV}$ and the radio loudness $R_{\rm UV}$ in this work, where $R_{\rm UV}$ of all objects are found to be larger than 100 (averagely higher than the 3CRR objects, see, e.g., \citealt{2021A&A...654A.141L}). A strong positive correlation is shown in Fig. \ref{f2}, which reads
\begin{equation}
R_{\rm UV}=(1.25\pm 0.33) \alpha_{\rm EUV}+0.52\pm 0.7,
\end{equation}
where the confidence level based on a Pearson test is about 99.7\%.

\section{DISCUSSION} \label{conclusionS}

The formation of MAD will affect both the inner region of thin accretion disk and the corona above disk, resulting on a deficit of EUV emissions. However, how MAD will change the dynamics of corona is unclear. Usually, corona is believed to be compact and small, of the order tens or less gravitational radii
($r_{\rm g} = GM/C^2$; e.g. \citealt{2014A&ARv..22...72U}), while of which the structure and composition is still under debate so far \citep{2018MNRAS.480.1819R}. Corona has been suggested to be similar with ADAF by some works (e.g., \citealt{2015ApJ...806..223L,2018MNRAS.477..210Q}). MAD can greatly increase the radiative efficiency of ADAF, possibly due to the higher optical depth (see \citealt{2019ApJ...887..167X} for details), which is qualitatively consistent with our results (see Fig. \ref{f1}).
Except for the X-ray excess, the $L_{\rm X}-L_{\rm UV}$ slope between RLAGNs with EUV deficit and RQAGNs is found to be well coincident (e.g., \citealt{2010A&A...512A..34L}). Our results suggest that the X-ray emission in RLAGNs with EUV deficit may originate from MAD and can be applied to constraint the theoretical model for MAD in future.

The other possibility for the X-ray excess in RLAGNs with EUV deficit is that the relativistic jet can also contribute to and even dominate the emission of X-ray. If so, both of the radio emission and the X-ray emission should come form jet, where the synchrotron radiation of electron corresponding to the former provides the seed photon required by the latter to process the inverse Compton scattering. Therefore, the radio luminosity $L_{\rm R}$ should be strongly correlated with the X-ray luminosity $L_{\rm X}$ (e.g., \citealt{2003MNRAS.343L..59H, 2003MNRAS.345.1057M}). We further explore the $L_{\rm R}-L_{\rm X}$ relationship in Fig. \ref{f3} to validate this assumption. However, no strong correlation is found between them, which indicate that the X-ray emission in RLAGNs with EUV deficit may be dominated by accretion disk. Furthermore, \citet{2021A&A...654A.141L} find a strong but steeper $L_{\rm R}-L_{\rm X}$ relationship, i.e., $\log(L_{\rm R}/L_{\rm Edd}) = (2.0 \pm 0.2) \log(L_{\rm X}/L_{\rm Edd}) +0.3 \pm 0.6$, in RLAGNs by compiling a sample from 3CRR catalog, which can also be qualitatively explained by disk-corona model (see their paper for details). For comparison, a much shallower relationship of $L_{\rm R}\sim L_{\rm X}$ is reported in low luminosity AGNs \citep{2003A&A...400.1007C, 2003MNRAS.345.1057M, 2016MNRAS.456.4377X, 2018MNRAS.481L..45L}. Therefore, in summary, the X-ray emission in RLAGNs with EUV deficit can reflect the effect of MAD on corona.  

Except for the excess of X-ray flux, we also report a strong relationship between $\alpha_{\rm EUV}$ and $R_{\rm UV}$ (Fig. \ref{f2}). We notice that a somewhat similar relationship of $\alpha_{\rm EUV}$ and $\eta_{\rm jet}$ ($=Q_{\rm jet}/L_{\rm bol}$) was given in P15. However, the jet power $Q_{\rm jet}$ in P15 is estimated with the radio emission at 151 MHz dominated by the lobe of jet, which however represents the emission from jets very long time ago, thus different from the present bolometric luminosity. We adopt the core emission at 5 GHz to calculate the radio loudness $R_{\rm UV}$ in this work. Both the core radio emission and UV radiation are from present activity, therefore, the radio loudness $R_{\rm UV}$ will be more reasonable indicator of jet efficiency, to be used to study the properties of MAD. In physics, increasing magnetic field strength will produce stronger radio luminosity $L_{\rm R}$ in jet \citep{1977MNRAS.179..433B}. Meanwhile, the stronger MAD due to increasing magnetic field will result on the higher level of EUV deficit (larger $\alpha_{\rm EUV}$, see P15). Therefore, the positive $\alpha_{\rm EUV}-R_{\rm UV}$ relationship in Fig. \ref{f2} can be qualitatively understood under the scenario of MAD.

\section{CONCLUSIONS}

In this letter, we compile a sample to investigate the X-ray properties in RLAGNs with EUV deficit, which are suggested to be powered by a MAD. The X-ray flux in RLAGNs with EUV deficit is found to be about 4.5 times larger than that in RQAGNs (Fig. \ref{f1}). However, the slope of $L_{\rm X}-L_{\rm UV}$ relationship is found to be well consistent with that in RQAGNs \citep{2010A&A...512A..34L}. Besides, the traditional strong $L_{\rm R}-L_{\rm X}$ relationship is absent in our sample. But a strong positive relationship between the radio-loudness $R_{\rm UV}$ and $\alpha_{\rm EUV}$ is discovered (see Fig. \ref{f2}). All these results can help to confine the theory for MAD. 

\section* {ACKNOWLEDGEMENTS}
We thank the reviewer for his/her very helpful comments. SLL thanks Dr. Fuguo Xie and Erlin Qiao for valuable discussion on MAD. This work is supported by the NSFC (grants 11773056, 11873073), Shanghai Pilot Program for Basic Research – Chinese Academy of Science, Shanghai Branch (JCYJ-SHFY-2021-013), and the science research grants from the China Manned Space Project with NO. CMSCSST-2021-A06. Minhua Zhou is supported by the Science and Technology Project funded by the Education Department of Jiangxi Province in China (Grant No. GJJ211733), and the Doctoral Scientific Research Foundation of Shangrao Normal University (Grant No. K6000449). This work has made extensive use of the NASA/IPAC Extragalactic Database (NED), which is operated by the Jet Propulsion Laboratory, California Institute of Technology, under contract with the National Aeronautics and Space Administration (NASA).

\bibliography{x-ray}{}
\bibliographystyle{aa}

\end{document}